
\input phyzzx
\def\abstract{\vskip\frontpageskip
\centerline{\bf {\titlestyle Abstract}} \vskip\headskip}

\def\chapter#1{\par \penalty-300 \vskip\chapterskip
\spacecheck\chapterminspace
\chapterreset \titlestyle{\bf \chapterlabel \ \ #1}
\nobreak\vskip\headskip \penalty 30000
\wlog{\string\chapter\ \chapterlabel}}
\def\e{\, {\rm e}}
%
%
\REFS\gouli{M. Goulian and M. Li,
{\it Phys.\ Rev.\ Lett.\ }{\bf 66} (1991) 2051.}
\REFSCON\dfku{P. Di Francesco and D. Kutasov,
{\it Phys.\ Lett.\ }{\bf 261B} (1991) 385;
Princeton preprint PUPT--1276 (September, 1991).}
\REFSCON\kita{Y. Kitazawa,
{\it Phys.\ Lett.\ }{\bf 265B} (1991) 262.}
\REFSCON\sataco{N. Sakai and Y. Tanii,
{\it Prog.\ Theor.\ Phys.\ }{\bf 86} (1991) 547.}
\REFSCON\dotsenko{V.S. Dotsenko,
Paris preprint PAR--LPTHE 91--18 (February, 1991).}
\REFSCON\taya{Y. Tanii and S. Yamaguchi,
{\it Mod.\ Phys.\ Lett.\ }{\bf A6} (1991) 2271.}
\REFSCON\albago{L. Alvarez-Gaum\'e, J.L.F. Barb\'on and C. G\'omez,
CERN preprint CERN--TH.6142/91 (June, 1991).}
\REFSCON\aodh{K. Aoki and E. D'Hoker,
UCLA preprint UCLA/91/TEP/32 (August, 1991).}
\REFSCON\polyakov{A.M. Polyakov,
{\it Phys.\ Lett.\ }{\bf 103B} (1981) 207.}
\REFSCON\ddk{F. David, {\it Mod.\ Phys.\ Lett.\ }{\bf A3} (1989) 1651;
J. Distler and H. Kawai,
{\it Nucl.\ Phys.\ }{\bf B321} (1989) 509.}
\REFSCON\sepo{N. Seiberg,
{\it Prog.\ Theor.\ Phys.\ Suppl.\ }{\bf 102} (1990) 319;
J. Polchinski,
in {\it Strings '90}, eds. R. Arnowitt {\it et al}.
(World Scientific, Singapore, 1991).}
\REFSCON\kitarev{Y. Kitazawa,
Harvard preprint HUTP--91/A034 (July, 1991).}
\REFSCON\polyakovself{A.M. Polyakov,
{\it Mod.\ Phys.\ Lett.\ }{\bf A6} (1991) 635.}
\REFSCON\grkl{D.J. Gross and I.R. Klebanov,
{\it Nucl.\ Phys.\ }{\bf B359} (1991) 3.}
\REFSCON\satafact{N. Sakai and Y. Tanii,
Tokyo Inst.\ of Tech.\ and Saitama preprint
TIT/HEP--173, STUPP--91--120 (August, 1991).}
\REFSCON\miya{D. Minic and Z. Yang, Texas preprint UTTG--23--91 (1991).}
\REFSCON\torus{N. Sakai and Y. Tanii,
{\it Int.\ J. of Mod.\ Phys.\ }{\bf A6} (1991) 2743;
M. Bershadsky and I.R. Klebanov,
{\it Phys.\ Rev.\ Lett.\ }{\bf 65} (1990) 3088;
{\it Nucl.\ Phys.\ }{\bf B360} (1991) 559;
I.M. Lichtzier and S.D. Odintsov,
{\it Mod.\ Phys.\ Lett.\ }{\bf A6} (1991) 1953.}
\REFSCON\open{B. Durhuus, H.B. Nielsen, P. Olesen and J.L. Petersen,
{\it Nucl.\ Phys.\ }{\bf B196} (1982) 498;
B. Durhuus, P. Olesen and J.L. Petersen,
{\it Nucl.\ Phys.\ }{\bf B198} (1982) 157; {\bf B201} (1982) 176;
O. Alvarez, {\it Nucl.\ Phys.\ }{\bf B216} (1983) 125.}
\REFSCON\gswp{M.B. Green, J.H. Schwarz and E. Witten,
{\it Superstring Theory} (Cambridge Univ.\ Press,
Cambridge, 1987) Vol.\ 1, 2;
A.M. Polyakov, {\it Gauge Fields and Strings}
(Harwood Academic Publishers, Chur, 1987).}
\REFSCON\bpz{A.A. Belavin, A.M. Polyakov and A.B. Zamolodchikov,
{\it Nucl.\ Phys.\ }{\bf B241} (1984) 333.}
\REFSCON\majumdar{P. Majumdar,
FNAL preprint FERMILAB--PUB--91/183--T (June, 1991).}
\REFSCON\mamose{E. Martinec, G. Moore and N. Seiberg,
{\it Phys.\ Lett.\ }{\bf 263B} (1991) 190.}
\REFSCON\df{V.S. Dotsenko and V.A. Fateev,
{\it Nucl.\ Phys.\ }{\bf B240} (1984) 312; {\bf B251} (1985) 691;
V.S. Dotsenko, {\it Advanced Studies in Pure Mathematics} {\bf 16}
(1988) 123.}
\REFSCON\gtw{A. Gupta, S.P. Trivedi and M.B. Wise,
{\it Nucl.\ Phys.\ }{\bf B340}  (1990) 475.}
\REFSCON\beku{M. Bershadsky and D. Kutasov,
Princeton and Harvard preprint PUPT--1283, HUTP--91/A047
(September, 1991).}
\refsend
%
%
\Pubnum={STUPP--91--121}
%
\titlepage
\title{\bf Two-Dimensional Quantum Gravity on a Disk}
\vskip 20mm
\author{Yoshiaki Tanii}
\andauthor{Shun-ichi Yamaguchi}
\vskip 10mm
\address{Physics Department, Saitama University \break
Urawa, Saitama 338, Japan}
\vskip 20mm
\abstract
We study a two-dimensional conformal field theory coupled to
quantum gravity on a disk.
Using the continuum Liouville field approach,
we compute three-point correlation functions of boundary operators.
The structure of momentum singularities is different from that of
correlation functions on a sphere and is more complicated.
We also compute four-point functions of boundary operators and
three-point functions of two boundary operators and one bulk operator.
\endpage
%
%
\chapter{Introduction}
Recently correlation functions of two-dimensional quantum gravity
have been computed \refmark{\gouli-\aodh} using the continuum
Liouville field approach\rlap.\refmark{\polyakov-\kitarev} They
exhibit a characteristic singularity structure, which was
analyzed in Refs.\ \polyakovself-\miya, \dfku.
Only correlation functions on a sphere were considered so far.
One may generalize these results to the case of
higher genus surfaces such as a torus\rlap.\refmark{\torus}
Another direction of generalizations is to consider surfaces
with boundaries.
Surfaces with boundaries naturally appear in string theories including
open strings in addition to closed strings\rlap.\refmark{\open, \gswp}
\par
The purpose of our paper is to study properties of two-dimensional
quantum gravity on a disk using the continuum Liouville field approach.
We will consider two-dimensional gravity coupled to a conformal
field theory \refmark{\bpz} realized by a single scalar field.
In the conformal gauge this model can be regarded
as a string theory in a two-dimensional target space.
\par
We first discuss the conformal gauge fixing on the disk
and obtain the effective Liouville action.
For surfaces without boundaries it was carried out in
Ref.\ \ddk. We generalize it to the case of surfaces with boundaries.
The conformal gauge fixing on surfaces with boundaries was also
discussed in Ref.\ \majumdar\ using the Dirichlet boundary
condition on the Liouville field. We use the Neumann boundary
condition on the Liouville field as well as on the matter field.
There are two kinds of physical operators on surfaces with
boundaries\rlap.\refmark{\mamose} Operators given by
two-dimensional integrals of local operators
in the bulk of the surfaces are called bulk operators.
On the other hand, operators given by line integrals of local operators
on the boundary are called boundary operators.
They correspond to closed string vertex operators and open string
vertex operators in string theories\rlap.\refmark{\gswp} Then we
compute correlation functions of such physical operators
and discuss their momentum singularities.
The three-point function of the boundary operators is mainly discussed.
The structure of the singularities is different from that
of a sphere topology and is more complicated.
We also compute the four-point function of the boundary operators and
the three-point function of two boundary operators and one bulk operator.
\par
In the next section we discuss the conformal gauge fixing.
In Sec.\ 3 the three-point function of the boundary operators are
obtained. Other correlation functions are discussed in Sec.\ 4.
An integral needed in Sec.\ 3 is computed in the Appendix.
\par
%
%
\chapter{Conformal Gauge Fixing}
We consider a scalar field $X$ coupled to gravity
on a two-dimensional manifold $M$ with a boundary $\partial M$
$$
S_{\rm M}›g,X!
= {1 \over 8\pi} \int\nolimits_{M} d^2 \xi \sqrt{g}
\left( g^{\alpha\beta} \partial_\alpha X \partial_\beta X
+ 2i\alpha_0 R X \right)
+{i \alpha_0 \over 2\pi} \int\nolimits_{\partial M} ds k X,
\eqn\maction
$$
where $\alpha_0$ is a real parameter and $R$ is the scalar curvature.
Without losing generality we choose $\alpha_0$ to be negative.
Using a real parameter $\tau$ the boundary
$\partial M$ is described by $\xi^\alpha = \xi^\alpha (\tau)$
and the line element on the boundary is given by
$$
d s = d \tau E, \quad
E = \left( g_{\alpha\beta} {d\xi^\alpha \over d\tau}
{d\xi^\beta \over d\tau} \right )^{1\over2}.
\eqn\lineelement
$$
The geodesic curvature $ k $ of the boundary is defined by
$$
t^\alpha \nabla_\alpha t^\beta = k n^\beta,
\eqn\geodesic
$$
where we have used unit tangent and normal vectors on the boundary
$$
t^\alpha = {d\xi^\alpha \over d\tau} E^{-1}, \qquad
n^\alpha = -{1 \over \sqrt{g}} \epsilon^{\alpha\beta}
g_{\beta\gamma} t^\gamma.
\eqn\tnvectors
$$
For a fixed metric the action \maction\ describes a conformal
field theory with a central charge \refmark{\df}
$$
c = 1 - 12 \alpha_0^2.
\eqn\mcc
$$
The boundary term in Eq.\ \maction\ is needed for a simple
transformation property of the action under the Weyl transformation.
The partition function of this system is
$$
Z = \int{{\cal D}_g g_{\alpha\beta} {\cal D}_g X \over V_{\rm gauge}}
\e^{-S_{\rm M}›g,X! - {\mu_0 \over \pi} \int_{M} d^2 \xi \sqrt{g}
-{\lambda_0 \over \pi} \int_{\partial M} ds},
\eqn\partition
$$
where $V_{\rm gauge}$ is the volume of the group
of two-dimensional general coordinate transformations.
Parameters $\mu_0$ and $\lambda_0$ are bare values of
bulk and boundary cosmological constants respectively.
The functional measures ${\cal D}_g g_{\alpha\beta}$ and
${\cal D}_g X$ are defined from the general coordinate invariant
metric on the functional space \refmark{\polyakov}
$$
\eqalign{
ºº \delta g_{\alpha\beta} ºº_g^2
& = \int d^2\xi \sqrt{g} \, g^{\alpha\gamma} g^{\beta\delta}
\delta g_{\alpha\beta} \delta g_{\gamma\delta}, \cr
ºº \delta X ºº_g^2
& = \int d^2\xi \sqrt{g} \, \delta X \delta X.
}\eqn\norms
$$
\par
{}From now on we consider the case of a disk.
We fix the general coordinate symmetry
by the conformal gauge condition \refmark{\polyakov}
$$
g_{\alpha\beta} = \e^{\alpha \phi} \hat g_{\alpha\beta},
\eqn\cgauge
$$
where $\hat g_{\alpha\beta}$ is a fixed reference metric and
$\alpha$ is a parameter to be fixed later.
An ansatz for the relation between ${\cal D}_g \phi$ and
${\cal D}_{\hat g} \phi$ was proposed in Ref.\ \ddk\ for
manifolds without boundaries.
We use a generalization of it to the case with a boundary.
According to such an ansatz the partition function
\partition\ becomes
$$
Z = \int {{\cal D}_{\hat g} \phi {\cal D}_{\hat g} X
\over V_{SL(2, {\bf R})}}
\Delta_{\rm FP}›\hat g!
\e^{-S_{\rm M}›\hat g, X! - S_{\rm L}›\hat g, \phi!},
\eqn\fixedpartition
$$
where $V_{SL(2, {\bf R})}$ is the volume of the gauge group generated by
the conformal Killing vectors on the disk, i.e. $SL(2, {\bf R})$
and $\Delta_{\rm FP}$ is the Faddeev-Popov determinant.
The effective Liouville action is given by
$$
\eqalign{
S_{\rm L}›\hat g, \phi!
= \ & {1 \over 8\pi} \int\nolimits_{M} d^2 \xi \sqrt{\hat g}
\left( \hat g^{\alpha\beta} \partial_\alpha \phi \partial_\beta \phi
- Q \hat R \phi + 8 \mu \e^{\alpha \phi} \right) \cr
& + {1 \over 4\pi} \int\nolimits_{\partial M} d \hat s
\left( -Q' \hat k \phi + 4\lambda \e^{{1 \over 2} \alpha' \phi} \right),
}\eqn\laction
$$
where $\mu$ and $\lambda$ are the renormalized cosmological constants.
\par
The parameters $Q, Q', \alpha$ and $\alpha'$ in Eq.\ \laction\
can be fixed by requiring that the theory does not depend on
the gauge choice $\hat g_{\alpha\beta}$.
In particular it should be invariant under a change
$$
\hat g_{\alpha\beta}(\xi)
\rightarrow \e^{\sigma(\xi)} \hat g_{\alpha\beta}(\xi)
\eqn\weyl
$$
for an arbitrary function $\sigma$.
As in Ref.\ \ddk\ we shall treat the cosmological terms
perturbatively. Thus we first examine the invariance of
the partition function \fixedpartition\ with $\mu = \lambda = 0$.
To obtain the transformation law of the Liouville part of
Eq.\ \fixedpartition\ we use a formula
$$
\eqalign{
S_{\rm L}›{\rm e}^\sigma \hat g ,\phi!
= \ & S_{\rm L} \bigl› \hat g, \phi - {Q \over 2}\sigma \bigr!
- {1 \over 8\pi}(Q-Q') \int\nolimits_{\partial M}
d \hat s \, \hat n^\alpha \partial_\alpha \sigma \phi \cr
& - {Q^2 \over 16\pi} \left›
\int\nolimits_{M} d^2 \xi \sqrt{\hat g} \left(
{1 \over 2} \hat g^{\alpha\beta}
\partial_\alpha \sigma \partial_\beta \sigma
+ \hat R \sigma \right)
+ {2Q' \over Q} \int\nolimits_{\partial M} d \hat s
\hat k \sigma \right!
}\eqn\lactionchange
$$
and the fact that the transformation law of the functional measure
${\cal D}_{\hat g} \phi$ under Eq.\ \weyl\ is the same as
that of a free scalar. If we choose $Q' = Q$, we obtain
$$
\eqalign{
\int {\cal D}_{{\rm e}^\sigma \hat g} \phi
& \e^{ -S_{\rm L}›{\rm e}^\sigma \hat g, \phi!}
= \int {\cal D}_{\hat g} \phi \, {\rm e}^{ -S_{\rm L}›\hat g, \phi!} \cr
& \times \ \exp \left›{c_{\rm L} \over 48\pi}
\left( \int\nolimits_{M} d^2 \xi \sqrt{\hat g} \left(
{1\over2} {\hat g}^{\alpha\beta} \partial_\alpha \sigma \partial_\beta \sigma
+ \hat R \sigma \right)
+ 2 \int\nolimits_{\partial M} d \hat s \hat k \sigma \right) \right!,
}\eqn\lpartition
$$
where $c_{\rm L} = 1 + 3Q^2$.
This relation shows that the Liouville sector is a conformal
field theory with the central charge $c_{\rm L}$.
The value of $c_{\rm L}$ is the same as in the case without boundaries.
Similarly, the matter and the Faddeev-Popov factors in
Eq.\ \fixedpartition\ can be shown to transform as
Eq.\ \lpartition\ with $c_{\rm L}$ replaced by $c$ and $-26$
respectively\rlap.\refmark{\open} Therefore the invariance of
the partition function \fixedpartition\ under the
change \weyl\ requires the vanishing of the total central charge
$c_{\rm L} + c - 26 = 0$, which determines
$$
Q = Q' = \sqrt{25 - c \over 3}.
\eqn\qq
$$
\par
Before we determine the parameters $\alpha, \alpha'$, let us discuss
several properties of the conformal field theory
described by the action
$$
S ›\hat g, X, \phi! = S_{\rm M} ›\hat g, X! + S_{\rm L} ›\hat g, \phi!
\eqn\totalaction
$$
with $\mu = \lambda= 0$.
It is convenient to choose a particular metric for $\hat g_{\alpha\beta}$.
We take a north half of the round sphere as the disk.
It has a positive constant curvature $\hat R$ and has a vanishing
geodesic curvature $\hat k = 0$ since the boundary is a geodesic.
The boundary condition on $X$ and $\phi$ can be determined
by the vanishing of boundary terms
in the variation of the action with respect to $X$ and $\phi$.
There are two possibilities: the Neumann boundary condition and the
Dirichlet one. Here we choose the former
$$
\hat n^\alpha \partial_\alpha X = 0, \quad
\hat n^\alpha \partial_\alpha \phi = 0 \quad {\rm on}\ \partial M.
\eqn\boundary
$$
The disk can be conformally mapped onto the upper-half complex
plane $\{ z \in {\bf C}º{\rm Im} z \ge 0 \}$. We can use
$\tau = x \equiv {\rm Re}z$ to parametrize the boundary
${\rm Im} z = 0$. On the upper-half plane
the fundamental operator product expansions (OPEs) consistent
with the boundary condition \boundary\ are
$$
\eqalign{
X(z, \bar z) X(w, \bar w) &
\sim - \ln ºz-wº^2 - \ln ºz-\bar wº^2, \cr
\phi(z, \bar z) \phi(w, \bar w) &
\sim - \ln ºz-wº^2 - \ln ºz-\bar wº^2.
}\eqn\ope
$$
The energy-momentum tensor is obtained from the variation of the action
with respect to the metric $\hat g_{\alpha\beta}$:
$$
\eqalign{
\delta S_{\rm L} ›\hat g, \phi!
= & {1\over8\pi}\int\nolimits_{M} d^2 \xi \sqrt{\hat g} \,
\delta \hat g^{\alpha\beta} \biggl›
\partial_\alpha \phi \partial_\beta \phi
-{1\over2} \hat g_{\alpha\beta} \hat g^{\gamma\delta}
\partial_\gamma \phi \partial_\delta \phi \cr
& + Q (\hat\nabla_\alpha \partial_\beta
- \hat g_{\alpha\beta} \hat\nabla^\gamma \partial_\gamma)
\phi \biggr!
- {Q \over 8\pi} \int\nolimits_{\partial M} d \hat s \,
\delta \hat g^{\alpha\beta}
\hat t_\alpha \hat t_\beta \hat n^\gamma \partial_\gamma \phi
}\eqn\lactionvar
$$
and a similar expression for $\delta S_{\rm M}$.
With the Neumann boundary condition \boundary\
the boundary term in Eq.\ \lactionvar\ vanishes.
The energy-momentum tensor is given by
$$
T(z) = - {1 \over 2} \partial_z X \partial_z X
+ i \alpha_0 \partial_z^2 X
- {1 \over 2} \partial_z \phi \partial_z \phi
- {1 \over 2} Q \partial_z^2 \phi,
\eqn\emtensor
$$
and the antiholomorphic component $\bar T(\bar z)$.
By the boundary condition \boundary\ they satisfy
$T = \bar T$ on the boundary $z = \bar z$.
Using Eq.\ \ope\ one can easily verify that the energy-momentum
tensor \emtensor\ satisfies the OPE of the Virasoro algebra
with the central charge $c_{\rm L} + c = 26$.
\par
The parameters $\alpha, \alpha'$ are fixed by requiring the invariance
of the cosmological term operators
$\int d^2 z \sqrt{\hat g} \e^{\alpha \phi(z)}$ and
$\int d x \hat E \e^{{1 \over 2} \alpha' \phi(x)}$ under the change \weyl.
Alternatively we can require that the integrands of these operators
have conformal weight one:
$$
\eqalign{
T (z) \e^{\alpha \phi(w)}
& \sim \partial_w \left( {1 \over z-w}
\e^{\alpha \phi(w)} \right), \cr
T (z) \e^{{1 \over 2} \alpha' \phi(x)}
& \sim \partial_x \left( {1 \over z-x}
\e^{{1 \over 2} \alpha' \phi(x)} \right).
}\eqn\weightone
$$
Using the energy-momentum tensor \emtensor\ and the OPEs \ope\
the requirements \weightone\ give conditions \refmark{\mamose}
$$
- {1 \over 2} \alpha^2 - {1 \over 2} Q \alpha = 1 =
- {1 \over 2} \alpha'^2 - {1 \over 2} Q \alpha'.
\eqn\alphacond
$$
Hence, we obtain
$$
\alpha = \alpha'
= - {1 \over 2\sqrt{3}} \left( \sqrt{25-c} - \sqrt{1-c} \right),
\eqn\alphasol
$$
where we have chosen a particular solution of
Eq.\ \alphacond\ following Ref.\ \sepo.
Thus we have determined all the parameters in the action \laction.
\par
We now turn to a discussion of physical operators.
There are two kinds of physical operators \refmark{\mamose}:
bulk operators (closed string vertex operators)
$$
O_{\rm c}(p) = \int d^2 z \sqrt{\hat g} \e^{ipX + \beta \phi}
\eqn\bulkop
$$
and boundary operators (open string vertex operators)
$$
O_{\rm o}(p)
= \int d x \hat E \e^{{1 \over 2}ipX + {1 \over 2}\beta \phi}.
\eqn\boundaryop
$$
The invariance of these operators under the change \weyl\ requires
that the integrands are primary fields of weight one and satisfy
the OPEs of the same form as Eq.\ \weightone.
We find that $p$ and $\beta$ must satisfy
$$
{1 \over 2} p^2 - \alpha_0 p
- {1 \over 2} \beta^2 -{1 \over 2} Q \beta = 1
\eqn\betacond
$$
for both of Eqs.\ \bulkop\ and \boundaryop\ and obtain
$$
\beta = - {Q \over 2} + ºp - \alpha_0º,
\eqn\betasol
$$
where we have again chosen a particular solution of Eq.\ \betacond\
following Ref.\ \sepo.
The cosmological term operators in Eq.\ \laction\ are a particular
case of the physical operators \bulkop, \boundaryop\ with $p = 0$.
\par
%
%
\chapter{Three-Point Functions of Boundary Operators}
In this section we compute the three-point correlation functions
of the boundary operators using the results in the previous section.
The correlation function of $N$ boundary operators \boundaryop\
and $M$ bulk operators \bulkop\ is given by a path integral
$$
\eqalign{
& \VEV{O_{\rm o}(p_1) \cdots O_{\rm o}(p_N)
O_{\rm c}(p_{N+1}) \cdots O_{\rm c}(p_{N+M})} \cr
& \qquad\quad
= \int {{\cal D}_{\hat g} X {\cal D}_{\hat g} \phi
\over V_{SL(2, {\bf R})}} \;
\e^{-S_{\rm M}›\hat g, X!-S_{\rm L}›\hat g, \phi!} \;
O_{\rm o}(p_1) \cdots O_{\rm c}(p_{N+M}).
}\eqn\pathintegral
$$
The action \laction\ has two cosmological constants $\mu$ and $\lambda$.
We consider the case $\mu = 0$ and $\lambda \not= 0$ for simplicity.
One may consider the case $\mu \not= 0$ and $\lambda = 0$ similarly.
When both of $\mu$ and $\lambda$ are non-zero, one can use a perturbation
expansion in one of them while treating the other exactly.
As in the case of correlation functions on a sphere \refmark{\gtw}
we can first integrate over the zero modes $X_0, \phi_0$
$(X = X_0 + \tilde X, \phi=\phi_0 + \tilde\phi)$
$$
\eqalign{
& \VEV{O_{\rm o}(p_1) \cdots O_{\rm c}(p_{N+M})} \cr
& \qquad\quad
= 2\pi \delta \left( \sum_{i=1}^N p_i + 2 \sum_{i=N+1}^{N+M} p_i
- 2 \alpha_0 \right)
{4 \Gamma (-s) \over º\alphaº}
\left( {\lambda \over \pi} \right)^s
\tilde A (p_1, \cdots, p_{N+M}),
}\eqn\atilde
$$
$$
\eqalign{
\tilde A (p_1, \cdots, p_{N+M}) =
& \; \int \prod_{i=1}^N \left› d x_i \hat E \right!
\int \prod_{i=N+1}^{N+M} \left› d^2 z_i \sqrt{\hat g} \right!
{1 \over V_{SL(2, {\bf R})}} \cr
& \times \VEV{ \prod_{i=1}^N \e^{{1 \over 2}i p_i \tilde X(x_i)}
\prod_{i=N+1}^{N+M} \e^{i p_i \tilde X(z_i)} }_{\tilde X} \cr
& \times \VEV{ \prod_{i=1}^N \e^{{1 \over 2} \beta_i \tilde\phi(x_i)}
\prod_{i=N+1}^{N+M} \e^{\beta_i \tilde\phi(z_i)}
\left( \int d x \hat E \e^{{1 \over 2}\alpha\tilde\phi} \right)^s
}_{\tilde\phi},
}\eqn\zeromodeint
$$
where we have defined
$$
s={1 \over º\alphaº} \left( Q + \sum_{i=1}^N \beta_i
+ 2 \sum_{i=N+1}^{N+M} \beta_i \right).
\eqn\es
$$
The expectation values in Eq.\ \zeromodeint\ are with respect to
the non-zero modes $\tilde X, \tilde\phi$, which have a free action
$$
S_0 ›\hat g, \tilde X, \tilde\phi! = {1 \over 8\pi} \int
d^2 z \sqrt{\hat g} \hat g^{\alpha\beta} \left(
\partial_\alpha \tilde X \partial_\beta \tilde X
+ \partial_\alpha \tilde \phi \partial_\beta \tilde \phi \right).
\eqn\freeaction
$$
When $s$ is a non-negative integer, we can evaluate the expectation
values. In this case one has to interpret a singular factor in
Eq.\ \atilde\ as \refmark{\torus, \grkl, \dfku}
$$
\Gamma (-s) \left( {\lambda \over \pi} \right)^s
\;\; \rightarrow \;\; (-1)^{s+1} {1 \over s|}
\left({\lambda \over \pi} \right)^s \ln \lambda.
\eqn\loglambda
$$
\par
Now we consider the case $N = 3, M = 0$,
i.e. the three-point function of the boundary operators.
By representing the gauge volume as
$$
V_{SL(2, {\bf R})}
= \int {dx_1 dx_2 dx_3 \over ºx_1-x_2º ºx_2-x_3º ºx_3-x_1º}
\eqn\slvolume
$$
and fixing the $SL(2, {\bf R})$ gauge symmetry
by choosing the positions of the boundary operators as
$x_1=0, x_2=1, x_3=\infty$, we obtain
$$
\eqalign{
\tilde A(p_1, p_2, p_3)
& = \int_{-\infty}^{\infty} \prod_{i=1}^s \left› dx_i \,
ºx_iº^{- \alpha \beta_1} º1-x_iº^{- \alpha \beta_2} \right!
\prod_{1 \leq i < j \leq s} ºx_i - x_jº^{-\alpha^2} \cr
& = s| \sum_{l,m,n=1 \atop l+m+n=s}^s I_{lmn}(a, b, \rho).
}\eqn\integralrep
$$
Here, we have introduced an integral
$$
\eqalign{
I_{lmn}(a,b,\rho)
& = {1 \over l| \, m| \, n|} \int_1^{\infty} \prod_{i=1}^l dx_i
\int_0^1 \prod_{i=l+1}^{l+m} dx_i
\int_{-\infty}^0 \prod_{i=l+m+1}^{l+m+n} dx_i \cr
& \quad \times \prod_{i=1}^{l+m+n} \left› ºx_iº^a º1-x_iº^b \right!
\prod_{1 \leq i < j \leq l+m+n} ºx_i-x_jº^{2\rho},
}\eqn\lmnint
$$
where
$$
a = - \alpha \beta_1, \quad b = - \alpha \beta_2,
\quad \rho = -{1 \over 2} \alpha^2.
\eqn\abrho
$$
The integral $I_{lmn}$ is a contribution from an integration region
where $l$ `cosmological terms' $\e^{\alpha\tilde\phi/2}$ are
inserted between operators 2 and 3, $m$ are inserted between
operators 1 and 2 and $n$ are inserted between operators 3 and 1.
It can be evaluated using a technique of Ref.\ \df.
We compute it in the Appendix. The result is
$$
\eqalign{
& I_{lmn}(a,b,\rho) = (-1)^s \cr
& \times \prod_{k=1}^l {\sin (\pi (a+(s-k)\rho))
\over \sin (\pi k\rho)}
\prod_{k=1}^m {\sin (\pi (c+(s-k)\rho)) \over \sin (\pi k\rho)}
\prod_{k=1}^n {\sin (\pi (b+(s-k)\rho)) \over \sin (\pi k\rho)} \cr
& \times \prod_{k=1}^s
{\Gamma(1+a+(k-1)\rho) \Gamma(1+b+(k-1)\rho) \Gamma(1+c+(k-1)\rho)
\over \Gamma(\rho) \Gamma(1-k\rho)},
}\eqn\intresult
$$
where $c= -2-a-b-2(s-1)\rho = -\alpha\beta_3$.
\par
By the momentum conservation and the definition of $s$
we have two conditions
$$
\eqalign{
p_1 + p_2 + p_3 & = - 2 \alpha - Q, \cr
ºp_1 - \alpha_0º + ºp_2 - \alpha_0º + ºp_3 - \alpha_0º
& = - s \alpha + {Q \over 2}.
}\eqn\emconservation
$$
There are two kinematical regions for the momenta
satisfying these conditions:
(i) $p_1, p_2 > \alpha_0, \; p_3 < \alpha_0$
and (ii) $p_1, p_2 < \alpha_0, \; p_3 > \alpha_0$.
As in the case of sphere topology \refmark{\dfku},
it is convenient to define
$$
m_i = {1 \over 2} \beta_i^2 - {1 \over 2} p_i^2.
\eqn\mm
$$
\par
Let us consider each case separately.
In the case (i), the momentum $p_3$ has a fixed value
$p_3 = {1 \over 2} (s-1) \alpha - {Q \over 2}$ by \emconservation\ and
only independent variable is $p_1$ (or $p_2$).
Eq.\ \intresult\ is simplified to
$$
\eqalign{
I_{lmn}(a,b,\rho) & = (-1)^{l+n} {\pi \over s|}
\left› {\pi \over \Gamma (1+\rho)} \right!^s
\prod_{k=1}^l {1 \over \sin (\pi k \rho)}
\prod_{k=1}^n {1 \over \sin (\pi k \rho)} \cr
& \quad \times {1 \over \Gamma (1-m_1) \Gamma (1-m_2)}
\prod_{k=0}^m {1 \over \sin (\pi (m_1 - (n+k)\rho))},
}\eqn\caseone
$$
where we have used the variables \mm\ given by
$m_1 = \rho + \alpha\beta_1, \;\; m_2 = 1 + s\rho - m_1,
\;\; m_3 = - s$ in this case. It has poles at
$$
m_1 = r + (n+k)\rho \qquad (k=0, \cdots, m; \;\; r \in {\bf Z}).
\eqn\poleone
$$
In the total amplitude \integralrep\ some of these poles may not
appear due to cancellations in the sum over $l, m, n$.
As an example, let us consider poles at
$m_1 = r + \rho \; (r \in {\bf Z})$ for $s=2$.
$I_{200}$ and $I_{002}$ are regular at $m_1 = r + \rho$.
Other $I_{lmn}$'s have poles there but they are canceled in combinations
$I_{020}+I_{101}$ and $I_{110}+I_{011}$. Therefore the total amplitude
\integralrep\ does not have a pole at $m_1 = r + \rho$.
However, except for such special cases, the poles \poleone\ in general
survive after the summation in Eq.\ \integralrep.
The amplitude \integralrep\ with Eq.\ \caseone\ has a quite different
form from that of a sphere topology\rlap.\refmark{\gouli-\aodh}
\par
Next we turn to the case (ii). The momentum $p_3$ has a fixed value
$p_3 = - {1 \over 2} (1+s) \alpha$ by Eq.\ \emconservation.
Substituting it into Eq.\ \intresult\ we find that $I_{lmn}$
vanishes when $m$ is non-zero.
Therefore we only have to consider the case $m=0$ in Eq.\ \intresult
$$
\eqalign{
I_{l 0 n}(a,b,\rho)
& = (-1)^l \left› {\pi \over \Gamma(1+\rho)} \right!^s
{1 \over \Gamma(1-s\rho)} \cr
& \quad \times \prod_{k=1}^l {1 \over \sin (\pi k \rho)}
\prod_{k=1}^n {1 \over \sin (\pi k \rho)}
\prod_{k=1}^s {1 \over 1-k-m_2},
}\eqn\casetwo
$$
where the variables \mm\ are given by
$m_1 = (1 - \alpha\beta_1) \rho^{-1}, \;\; m_2 = 1 - s - m_1,
\;\; m_3 = s \rho$. It has a finite number of poles at
$$
m_2 = 1 - k \qquad (k = 1, 2, \cdots, s).
\eqn\poletwo
$$
The pole structure \poletwo\ is independent of $l, m, n$.
As a consequence the sum in Eq.\ \integralrep\ identically vanishes
when $s$ is an odd integer. However, for an even integer $s$ the
total amplitude \integralrep\ does not vanish in general.
This is in contrast to the case of sphere topology,
in which correlation functions in the kinematical region (ii)
identically vanish\rlap.\refmark{\dfku}
\par
%
%
\chapter{Other Correlation Functions}
Here we shall discuss some of other correlation functions briefly.
First let us compute the four-point correlation function
of the boundary operators \boundaryop.
It is given by Eq.\ \zeromodeint\ with $N = 4, M = 0$.
We only consider the case $s=0$.
After fixing the $SL(2, {\bf R})$ symmetry by choosing
$x_1=0, x_3=1,x_4=\infty$, the amplitude \zeromodeint\ becomes
$$
\eqalign{
\tilde A(p_1, p_2, p_3, p_4)
& = \int_{-\infty}^{\infty}
dx_2 ºx_2º^{p_1 \cdot p_2} º1-x_2º^{p_2 \cdot p_3} \cr
& = {\Gamma(1+p_1 \cdot p_2) \Gamma(1+p_2 \cdot p_3)
\over \Gamma(-p_2 \cdot p_4)} +
{\Gamma(1+p_2 \cdot p_4) \Gamma(1+p_1 \cdot p_2)
\over \Gamma(-p_2 \cdot p_3)} \cr
& \quad + {\Gamma(1+p_2 \cdot p_3) \Gamma(1+p_2 \cdot p_4)
\over \Gamma(-p_1 \cdot p_2)},
}\eqn\fouramplitude
$$
where $p_i \cdot p_j = p_i p_j -  \beta_i \beta_j$.
Three terms in \fouramplitude\ come from integration regions with
different orderings of the boundary operators $1, 2, 3$ and $4$.
The first, the second and third terms correspond to the orderings
(1, 2, 3, 4), (1, 3, 4, 2) and (1, 4, 2, 3) respectively. From
the momentum conservation and the definition of $s$,
three kinematical regions are possible:
(i) $p_1, p_2, p_3 > \alpha_0, \; p_4 < \alpha_0$,
(ii) $p_1, p_2 > \alpha_0, \; p_3, p_4 < \alpha_0$ and
(iii) $p_1, p_2, p_3 < \alpha_0, \; p_4 > \alpha_0$.
Using the variables \mm, the amplitude \fouramplitude\ in
the cases (i) and (iii) can be written as
$$
\tilde A(p_1, p_2, p_3, p_4)
= {\Gamma(m_3) \Gamma(m_1) \over \Gamma(1-m_2)}
+{\Gamma(m_2) \Gamma(m_3) \over \Gamma(1-m_1)}
+{\Gamma(m_1) \Gamma(m_2) \over \Gamma(1-m_3)}.
\eqn\caseonetwo
$$
In these cases the momentum $p_4$ has a fixed value due to
the conditions similar to \emconservation. The origin of the poles
in Eq.\ \caseonetwo\ is short distance singularities which arise
when the operator 4 approaches one of the neighbor operators.
On the other hand, in the case (ii) the third term in
Eq.\ \fouramplitude\ corresponding to the ordering (1, 4, 2, 3)
vanishes. Each of other two terms is non-zero but the sum vanishes.
Therefore the total amplitude is zero in this case.
\par
Next, as an example of correlation functions including both of
the bulk operators \bulkop\ and the boundary operators \boundaryop,
we consider the correlation function of two boundary operators
and one bulk operator.
Again we consider the case $s=0$.
The amplitude \zeromodeint\ with $N = 2, M = 1$ is
$$
\tilde A(p_1, p_2, p_3)
= \int {dx_1 dx_2 d^2z_3 \over V_{SL(2,{\bf R})}} \;
ºx_1 - x_2º^{p_1 \cdot p_2} ºz_3 - x_1º^{2p_1 \cdot p_3}
ºz_3 - x_2º^{2p_2 \cdot p_3} ºz_3 - \bar z_3º^{2 p_3 \cdot p_3},
\eqn\ooc
$$
where the last factor of the integrand is due to self-contractions
of $\tilde X(z_3), \tilde \phi (z_3)$ in the bulk
operator\rlap.\refmark{\gswp} Using a representation
of the gauge volume
$$
V_{SL(2,{\bf R})}
= \int {d^2 z_3 \, d x_2 \over ºz_3 -x_2º^2 \, {\rm Im}z_3}
\eqn\slvolumetwo
$$
and fixing $x_2 = 0, z_3 = i$, we obtain
$$
\tilde A(p_1, p_2, p_3)
= \sqrt{\pi} \; 2^{2 p_3 \cdot p_3} \;
{\Gamma({1 \over 2}p_1 \cdot p_2+{1 \over 2})
\over \Gamma({1 \over 2}p_1 \cdot p_2+1)}.
\eqn\oocresult
$$ From
the momentum conservation and the definition of $s$,
there are four cases of kinematical regions:
(i) $p_1, p_2 > \alpha_0, p_3 < \alpha_0 $,
(ii) $p_1 < \alpha_0, p_2, p_3 > \alpha_0$,
(iii) $p_1, p_2 < \alpha_0, p_3 > \alpha_0$
and (iv) $ p_1 > \alpha_0, p_2, p_3 < \alpha_0$.
In the cases (i) and (iii) the argument of the Gamma function
in the numerator is constant and there is no singularity in momenta.
In the cases (ii) and (iv) the argument depends on the momentum
$p_2$ (or $p_3$) and the amplitude \oocresult\ has pole singularities.
Thus the singularities arise only when $p_1 - \alpha_0$ and
$p_2 - \alpha_0$ have opposite signs.
The origin of these singularities is short distance singularities of
two boundary operators approaching each other.
\par
\vskip 5mm
\hskip -7mm{\it Note added\/}:
After completion of this work, we received
a preprint by Bershadsky and Kutasov \refmark{\beku}
discussing two-dimensional open string theory.
They give results of the general $N$-point correlation functions
of the boundary operators for $s=0$. Our results
\caseone, \casetwo\ and \caseonetwo\ are consistent with theirs.
\par
\vskip 5mm
%
%
\titlestyle{\bf Appendix}
In this Appendix we sketch an evaluation of the integral \lmnint.
We closely follow a method used in Ref.\ \df\ to evaluate
a similar integral.
For more details of the arguments see Ref.\ \df.
In order to evaluate the integral \lmnint,
it is convenient to introduce a complex contour integral
$$
\eqalign{
J_{lmn}(a,b,\rho)
& = \prod_{i=1}^{l+m+n} \left› \int\nolimits_{C_i} dz_i \right!
\prod_{i=1}^{l} \left›z_i^a (z_i-1)^b \right!
\prod_{i=l+1}^{l+m} \left›z_i^a (1-z_i)^b \right! \cr
& \quad \times \prod_{i=l+m+1}^{l+m+n} \left›(-z_i)^a (1-z_i)^b \right!
\prod_{1 \leq i < j \leq l+m+n} (z_i -z_j)^{2\rho}.
}\eqno{({\rm A}.1)}
$$
The contours $C_1, \cdots, C_l$ connect points 1 and $\infty$,
$C_{l+1}, \cdots, C_{l+m}$ connect 0 and 1, and
$C_{l+m+1}, \cdots, C_{l+m+n}$ connect $-\infty$ and 0
in the complex plane. Furthermore, in each of the three groups
the contour $C_i$ lies above the contour $C_j$ if $i > j$.
The phase of the integrand is defined to be zero
when all the integration variables are on the real axis
and $x_i > x_j$ for $i < j$.
It can be shown that the above integral is related to
Eq.\ \lmnint\ as
$$
\eqalign{
J_{lmn}(a,b,\rho)
& = \prod_{k=1}^{l} \left› {\rm e}^{-i\pi(k-1)\rho}
{\sin (\pi k \rho) \over \sin (\pi \rho)} \right!
\prod_{k=1}^{m} \left› {\rm e}^{-i\pi(k-1)\rho}
{\sin  (\pi k \rho) \over \sin (\pi \rho)} \right! \cr
& \quad \times \prod_{k=1}^{n} \left› {\rm e}^{-i\pi(k-1)\rho}
{\sin (\pi k \rho) \over \sin (\pi \rho)} \right!
I_{lmn}(a,b,\rho).
}\eqno{({\rm A}.2)}
$$
\par
Consider an integral of the form (A.1) with the contour $C_{l+m+n}$
replaced by a contour consisting of three parts:
a contour connecting $-\infty$ and $0$, a contour connecting $0$ and $1$,
and a contour connecting $1$ and $\infty$. The new contour lies above
all other contours $C_1, \cdots, C_{l+m+n-1}$.
It can be closed at infinity and the integral vanishes. From
the vanishing of the integral we obtain a recursion relation
$$
J_{lmn}
+ {\rm e}^{-i\pi (a + 2(n-1)\rho)}J_{l\,m+1\,n-1}
+ {\rm e}^{-i\pi (a + b + 2(n+m-1)\rho)}J_{l+1\,m\,n-1}
= 0.
\eqno{({\rm A}.3)}
$$
Similarly, consider an integral (A.1) with the contour $C_{l+m+1}$
replaced by a contour consisting of the above mentioned three parts
but now lying below all other contours $C_i \; (i \not= l+m+1)$. From
the vanishing of such an integral we obtain
$$
J_{lmn}
+ {\rm e}^{i\pi (a + 2m\rho)} J_{l\,m+1\,n-1}
+ {\rm e}^{i\pi (a + b + 2(m+l)\rho)} J_{l+1\,m\,n-1}
=0.
\eqno{({\rm A}.4)}
$$
Using Eqs.\ (A.3) and (A.4) we obtain
$$
\eqalign{
& J_{lmn} = (-1)^{l+n} {\rm e}^{i\pi (lm+mn+nl) \rho} \,
J_{0 \, l+m+n \, 0} \cr
& \;\; \times {\prod_{k=1}^l \sin›\pi (a + (m+n+k-1)\rho)!
\prod_{k=1}^n \sin›\pi (b + (m+l+k-1)\rho)! \over
\prod_{k=1}^{l+n} \sin ›\pi (a + b + (2m+n+l+k-2)\rho)!}.
}\eqno{({\rm A}.5)}
$$
Therefore the evaluation of the integral \lmnint\ has reduced to the
evaluation of the integral of the form $I_{0 n 0}$.
Such an integral was evaluated in Ref.\ \df
$$
I_{0 n 0} = \prod_{k=1}^n {\Gamma (k\rho) \over \Gamma (\rho)}
\prod_{k=0}^{n-1} {\Gamma (1+a+k\rho) \Gamma (1+b+k\rho) \over
\Gamma (2+a+b+(n-1+k)\rho)}.
\eqno{({\rm A}.6)}
$$ From Eqs.\ (A.2), (A.5) and (A.6)
we obtain the result in Eq.\ \intresult.
\endpage
\par \penalty-400 \vskip\chapterskip
\spacecheck\referenceminspace \immediate\closeout\referencewrite
\referenceopenfalse
\line{\hfil {\titlestyle {\bf References}} \hfil}\vskip\headskip
\input reference.aux
\end